\documentstyle[preprint,tighten,aps]{revtex}
\begin{document}

\draft
\title{Density and Graviton Perturbations in the Cosmic Microwave 
Background\footnote{Research partially funded by 
 NATO Collaborative Research Grant CRG no. 920129}}
\author{A. B. Henriques, L. E. Mendes}
\address{Departamento de Fisica,Instituto Superior Tecnico,
1096 Lisbon, Portugal}
\author{R. G. Moorhouse}
\address{Department of Physics and Astronomy,University of Glasgow,
Glasgow G12 8QQ, U.K.}
\date{\today}
\maketitle
\begin{abstract}
We evaluate and compare the gravitational wave and density
perturbation contributions to the cosmic microwave background
radiation, on the basis of the same power law inflationary model. 
The inflation to radiation transition is treated in this paper as
instantaneous, but a model is constructed to allow for a smooth
transition from the radiation to the matter dominated eras. The
equations are numerically investigated and integrated, without any
basic approximations being made. Use is made of the synchronous 
gauge, with appropriate gauge invariant variables, thus eliminating 
any confusion arising from unphysical gauge modes.  We find a non- 
negligible gravitational wave contribution, which becomes dominant 
for a power law expansion with exponent $q < 13$.  We also explore 
the dependence of our results with the main characteristic of the
transition region, its length.
\end{abstract}

\narrowtext
\section{Introduction} \label{intro}In the inflationary scenario 
graviton (metric tensor) and density (metric scalar) perturbations 
arise from quantum fluctuations \cite{PZSG} early in the inflationary
era. In many models the subsequent development of these separate
perturbations is governed by the same parameters of the given
model. Thus the absolute magnitude of the perturbations presently
observed in the CMBR and through other phenomena is predicted in terms
of these parameters so that, among other things, a comparison of the
metric tensor and metric scalar contributions can be made.  Models
with power law inflation are of this type. It is the purpose of this
paper to evaluate and compare the two contributions for such
models. In doing so we shall make the instantaneous approximation for
the transition from the inflationary era to the radiation era, since
it holds with good accuracy over the range of metric perturbation 
wavelengths of most physical interest\cite{Der,Cal}; we do not make 
that approximation for the transition from the radiation era to 
the matter era.

We take power law inflation to be driven by a single canonically
normalized scalar field with Einstein gravity. The assumption of a
single scalar field is more widely encompassing than it seems; almost
any current extended gravity theory, such as a higher order or
scalar-tensor theory can be rewritten as Einstein gravity using a
conformal transformation. Then, moreover, slow roll inflation
solutions (giving the standard calculations of density perturbations)
can be expanded about power law inflation solutions \cite{GLLS}.

We use a synchronous gauge for the formalism of the computing
programs, while extracting gauge invariant quantities \cite{MFB,Der}
for making physical comparisons and for delicate procedures of
conveying information from one cosmic era to another. Thus we
eliminate any confusion from unphysical synchronous gauge modes from
our results.

\section{The Inflationary Era}
\label{infera}We calculate with perturbations in a synchronous gauge 
so that the metric is
\begin{equation}\label{eq.metric}
ds^2 = -a^{2}d\tau^{2} + a^{2}(\delta_{ij} + h_{ij})dx^{i}dx^{j}
\end{equation}
In the inflationary era , with scalar field $\phi$, the
energy-momentum tensor is \begin{equation}\label{EQ.Tmunu}
T_{\mu\nu}=\phi_{,\mu}\phi_{,\nu}-g_{\mu\nu}
\lbrack\frac{1}{2}g^{\alpha\beta}\phi_{,\alpha}\phi_{,\beta}+
V(\phi)\rbrack
\end{equation}

\subsection{Metric Scalar Perturbations}
\label{infera1}
In this section we shall first treat the basic formalism and then 
specialise to the case of power-law inflation.
 
We adopt the notations of Grishchuk \cite{Gri} so that for density  
perturbations with wave number ${\bf k}$
\begin{equation}\label{eq.hij}
h_{ij} = h(\tau)Q\delta_{ij} + h_{l}(\tau)k^{-2}Q_{,i,j} 
\end{equation} 
where $Q = \exp(i{\bf k.x})$.

If $\rho_{1}$ and $p_{1}$ are the perturbations to the energy and 
pressure densities then the Einstein equations yield, where 
$\kappa = 8\pi G$,
\begin{equation}
a^{2}\kappa \rho_{1}  =  3\alpha h' + k^{2} h - \alpha h'_{l},
\label{EQ.ro}
\end{equation}
\begin{equation}
a^{2}\kappa p_{1}  =  -h'' - 2\alpha h',\label{EQ.pe}
\end{equation}
and also with the hypothesis of non-diagonal space-space 
components in the energy-momentum tensor, as with a scalar field 
or with a perfect fluid
\begin{equation}\label{EQ.hl}
0 = h''_{l} + 2\alpha h'_{l} - k^{2} h
\end{equation}
where  $ \alpha \equiv a'/a $. 

These equations being valid with an energy-momentum tensor of a 
scalar field or a perfect fluid (or a mixture) are applicable through 
all the model eras we shall consider from inflation to the present.

We now consider the inflationary era case with the
energy-momentum tensor given by Eq.({\ref{EQ.Tmunu}}). Then if the
unperturbed value of $\phi$ is $\phi_{0}$ and the perturbation to it
is $\phi_{1}$ the Einstein equations yield
\begin{equation}\label{EQ.phi}
\kappa\phi_{1} = h'/\phi'_{0}
\end{equation}
and also \cite{Gri}
\begin{equation}
\mu'' + 
\mu\lbrack k^{2}-(a\sqrt{\gamma})''/(a\sqrt{\gamma})\rbrack=0
\label{eq.mu''}
\end{equation}
\begin{equation}
\mu  \equiv  \frac{a}{\alpha\sqrt{\gamma}}(h' + \alpha\gamma h),
\label{EQ.mu}
\end{equation}
\begin{equation}
\gamma  \equiv  1 - \alpha'/\alpha^{2}. \label{eq.gamma}
\end{equation}
For the particular case of power law inflation the scale factor 
\begin{equation}\label{EQ.ap}
a(\tau) \propto (\tau_{i} - \tau)^{p}
\end{equation}
where $p ( <-1 ),\tau_{i}$ are constants so that $\gamma = 
(1 + p)/p = {\rm constant}$. Our model is one of distinctly 
power-law, rather than exponential, inflation; that is we do 
not approach the de Sitter limit, $p \to -1$ and $\gamma \to 0$.
If $t$ is the appropriate cosmic time for the inflation era then
$a(t) \propto t^{q}$ where $q = 1/\gamma$. An adequate amount of 
inflation requires $q \ge 10$; for larger $q$ we limit ourselves 
to values where expressions such as Eq.(\ref{EQ.hy}) below are 
extremely good approximations and this is certainly true up to 
$q = 100$.   

The solution for $h$ is given in terms of that for $\mu$ by 
Eq.(\ref{EQ.mu}) as
\begin{equation}\label{EQ.h}
h = \frac{\alpha}{a} \lbrace \sqrt{\gamma} \int^{\tau}
\mu(k,\tau')d\tau' + C_{i} \rbrace ,
\end{equation}
where $C_{i}$ is an integration constant subsuming, when $\gamma$ is  
constant, the lower limit of the integration. 
$h = \frac{\alpha}{a}C_{i}$ is a solution of $\mu = 0$  
corresponding to an unphysical mode of the 
synchronous gauge \cite{Gri} and we drop this term since, giving 
zero contribution to the gauge invariant variables, it does not 
carry information through into the radiation era , 
as related in the next section.

All the above development was non-quantum mechanical. We denote the 
corresponding quantum field theory quantities by a tilde:
\begin{equation}\label{eq.htil}
\tilde{h} = \frac{\alpha}{a} \sqrt{\gamma} \int^{\tau} 
\tilde{\mu}(k,\tau')d\tau'   
\end{equation}
\begin{equation}\label{EQ.mutil}  \tilde{\mu} = 
N\int \frac{d^{3}k}{(2\pi)^{\frac{3}{2}}\sqrt{2k}} 
\lbrack c_{k}\mu_{1}(k,\tau)\exp(i{\bf k.x}) 
+ h.c. \rbrack
\end{equation}
where $c_{k}$ is a quantum annihilation operator, $\lbrack
c_{k},c_{k'}^{\dag}\rbrack=\delta^{3}({\bf k-k'})$,
and $\mu_{1}(y), y = |k(\tau_{i} - \tau)|$, is that solution of the
Bessel equation (\ref{eq.mu''}) (with $a,\gamma$ specified by 
power-law inflation as above) such that
\begin{equation}\label{eq.limits}
\mu_{1}(y) \to e^{-iy}, y \to \infty 
\end{equation}
thus corresponding for large $k$ to the usual mode function of 
quantum mechanical plane waves.

$N$ is a normalization factor whose determination gives the 
absolute magnitude of the observed density perturbations (in terms 
of the model parameters) given the assumption that these come from 
a primordial vacuum with zero quantum occupation number. 
The determination of $N$ is as follows. 
 
Using the methods of ref.\cite{MFB} and Eqs.(\ref{EQ.phi},
\ref{EQ.hl'}) the gauge invariant scalar field, 
$\phi_{1}^{gi} \equiv \phi_{1} - \frac{1}{2}\phi'_{0}k^{-2}h'_{l}$ 
is given by the gauge invariant version of Eq.(\ref{EQ.phi}) as 
\begin{equation}\label{eq.phigi}
\kappa \phi_{1}^{gi} = \frac{\alpha}{a} \sqrt{\gamma} \lbrack -\mu +
\alpha \gamma k^{-2} \lbrace -\mu' + \mu(\alpha + \gamma'/2\gamma )
\rbrace \rbrack /\phi'_{0}
\end{equation}
where, for power law inflation, $\sqrt{\kappa} \phi'_{0} = -\alpha
 \sqrt{2\gamma}$. So in the limit $k \to \infty$ it follows that 
$\sqrt{2\kappa} \phi_{1}^{gi} \to \tilde{\mu}/a = 
N\tilde{\mu_{1}}/a$. This yields the result
\begin{equation}\label{eq.N}
N = \sqrt{2\kappa} = \sqrt{16\pi G}
\end{equation} 
since then
\begin{equation}\label{eq.phitilgi}
\phi_{1}^{gi} \to a^{-1}\int \frac{d^{3}k}{(2\pi)^{\frac{3}{2}}
\sqrt{2k}} \lbrack c_{k}\mu_{1}(k,\tau)\exp(i{\bf k.x})
 + h.c. \rbrack
\end{equation}
which is the appropriate quantization for a scalar field with 
the normalization (\ref{EQ.Tmunu}). This is the same normalizing 
factor as that found by 
Grishchuk \cite{Gri} using a similar argument. Also it is the same 
normalizing factor as that appropriate for the metric tensor case
\cite{Whi,Gri2} given below.  

We can now proceed with the evaluation of the metric scalar 
components at the end of the power-law inflation. The exact expression 
for the solution $\mu_{1}$ is, with $n = \frac{1}{2} - p$, 
\begin{equation}\label{eq.mu1}
\mu_{1}(y) = \sqrt{\frac{\pi y}{2}}(J_{n} - iY_{n})
\exp\lbrack-i(\frac{1}{2}n\pi + \frac{1}{4}\pi)\rbrack
\end{equation}    
As previously stated we shall assume a sudden transition at the 
end of inflation and the beginning of the radiation era; 
variables at that interface we denote by $\tau =
 \tau_{2}, a = a_{2}, H = H_{2} = -p((\tau_{i} -\tau_{2})a_{2})^{-1} 
 = (\tau_{2}a_{2})^{-1}$. The values of $k$ of
 interest are of the order of magnitude (the relevant meaning here is
 being within a few factors of ten) of
\begin{equation}\label{EQ.k1} 
k_{1} \equiv a'_{1}/a_{1} = a_{1}H_{1}
\end{equation}
where $a_{1},H_{1}$ denote the values of the scale factor and Hubble
at the time, $\tau_{1}$, when the matter era begins. Thus for such
values of $k$ the values of $y = |k(\tau_{i} - \tau_{2})|$ at the end
of the inflationary era, beginning of the radiation era, 
are of order $10^{-n} , n > 10$. So
then we can expand $\mu$ in a power series with leading terms
\begin{equation}
\mu_{1}(y) = M(p)\lbrack y^{p}-\frac{y^{p+2}}{2(2p+1)}.....\rbrack,
\label{eq.mu1y}
\end{equation}
\begin{equation}
M(p) \equiv -2^{-p}\Gamma(\frac{1}{2}-p)\exp(-i(1-p)/2) /\sqrt{\pi}.
\label{eq.Mp}
\end{equation}
The corresponding expansion of $h$ from Eq.(\ref{EQ.h}), with
 $C_{i}=0$, is 
\begin{equation}\label{EQ.hy} h = (\sqrt{\gamma}a)^{-1}
M(p)\lbrack y^{p} - \frac{(p+1)y^{p+2}}{2(2p+1)(p+3)}.....\rbrack
\end{equation}
We shall need these expressions at the interface of the inflationary
and radiation eras, $\tau = \tau_{2}$.

\subsection{Metric Tensor Perturbations}
\label{infera2}

The mode function $\mu/a$ for gravitons is given by 
 \begin{equation}\label{EQ.mugrav} 
\mu'' + \mu(k^{2} - a''/a) = 0 .
\end{equation}
This holds for any cosmic era, whatever may be the dynamics
 responsible for the particular form of $a(\tau)$. The quantum field
 theory expression for tensor perturbations is \cite{Whi,Gri2}
\widetext
\begin{equation}\label{EQ.htilgrav} 
\tilde{h}_{ij} = \sqrt{16\pi G}\sum_{\lambda =1}^{2}\int 
\frac{d^{3}k}{(2\pi)^{\frac{3}{2}}\sqrt{2k}a(\tau)} 
\lbrack a_{\lambda k}\epsilon_{ij}^{\lambda}(k)
\mu(k,\tau)\exp(i{\bf k.x}) + h.c. \rbrack,
\end{equation}
\narrowtext
where $a_{\lambda k}$ is the annihilation operator for the graviton
with polarization $\lambda$ and wave number ${\bf k}$,
\begin{equation}\label{eq.mem1}
\mu = \mu_{1}
\end{equation}
and the polarization tensor satisfies 
$\sum_{i,j}\epsilon_{ij}^{\lambda}(k) \epsilon_{ij}^{\lambda'}(k) = 
2\delta_{\lambda \lambda'}$.

\section{Into and Through the Radiation Era}
\label{itre}
We take the instantaneous approximation to the transition from
inflation to radiation. This is well justified because for the values
of $k$ of interest (discussed above in \ref{infera1}) and with a time
of transition of order $\tau_{i} - \tau_{2} = p/a_{2}H_{2}$, then the
parameter measuring the suddenness of the transition is, from
Eq.(\ref{EQ.k1}), of order $k_{1}(\tau_{i} - \tau_{2}) =
pa_{1}H_{1}/a_{2}H_{2} \approx a_{2}/a_{1}$, and is thus exceedingly
small. We use the method of Deruelle and Mukhanov
\cite{Der} to match the end of inflation to the beginning of radiation. 

Our convention is that the conformal time $\tau$ is continuous from 
inflation through the matter era, and that in the pure radiation era 
the scale factor  
\begin{equation}\label{eq.ataurad} 
a \propto \tau
\end{equation}
It follows that in the radiation era, $\tau = (aH)^{-1}$. In the
 inflation era the scale factor is proportional to $(\tau_{i} -
 \tau)^{p}$ (Eq.\ref{EQ.ap}) and in the pure matter era to $(\tau -
 \tau_{m})^{2}$ where $\tau_{i},\tau_{m}$ and the constants of
 proportionality are determined from the continuity of $a,a'$, with
 the convention $a({\rm present}) = 1$.

\subsection{Metric Scalar Perturbations}
\label{itre1}
In the radiation era, treated as a relativistic perfect fluid phase
\begin{equation}
\nu'' + \frac{1}{3}k^{2}\nu=0 ,
 \nu \equiv \frac{a}{\alpha}(h' + \alpha\gamma h) ,
\label{EQ.nu}
\end{equation}
where 
\begin{equation}
h  = \frac{a}{\alpha}\int^{\tau} \nu d\tau ,
 \gamma \equiv 1 - \alpha'/\alpha^{2} = 2 .\label{eq.hnu}
\end{equation}
The mode function $\mu$ of the inflationary era is replaced by $\nu$. 
We write the solution of Eq.(\ref{EQ.nu}) as
\begin{equation}\label{EQ.nusoln} 
\nu = B_{+}\cos (k(\tau - \tau_{2})/\sqrt{3}) - 
B_{-}\sin (k(\tau - \tau_{2})/\sqrt{3}).
\end{equation}

The gauge invariant function $\Phi$ \cite{MFB} is given by   
\begin{equation}\label{EQ.Phi} 
\Phi = h - (\alpha/k^{2})h'_{l},
\end{equation}
and using the Einstein equations, Eqs.(\ref{EQ.ro},\ref{EQ.pe}), with 
$p_{1} = \rho_{1}/3$, and Eq.(\ref{EQ.nu}) 
\begin{equation} 
\Phi = -(\nu'/\alpha - \nu)/a\epsilon^{2},
\epsilon = k/(\alpha\sqrt{3}).\label{EQ.Phieps}
\end{equation}
The matching conditions found in ref.\cite{Der}, by deduction from 
the Lichnerowicz conditions \cite{Lic}, imply that $\Phi$ and
the expression
\begin{equation}\label{eq.Gamma} 
\Gamma \equiv (\Phi'/\alpha + \Phi + \epsilon^{2}\Phi)/\gamma,
\end{equation}
should both be continuous on the surface $\tau = \tau_{2}$. On the
radiation side of the surface, $\tau = \tau_{2+}$:
\begin{equation} 
\Phi = (\epsilon_{2}B_{-} + B_{+})/(a_{2}\epsilon_{2}^{2}),
\end{equation}
\begin{equation}
\Gamma = -\lbrack B_{+}(1-\epsilon_{2}^{2}) + \epsilon_{2} 
B_{-}(1-\frac{1}{2}\epsilon_{2}^{2})\rbrack /(a_{2}\epsilon_{2}^{2}).
\label{eq.PhiGamma}
\end{equation}
We now need the values of $\Phi$ and $\Gamma$ on the inflation side of
the surface, $\tau = \tau_{2-}$. In the inflationary era the Einstein
equations yield
\begin{equation}\label{EQ.hl'} 
h'_{l} = \lbrack h'' + h'(\alpha-2\alpha'/\alpha-\gamma'/\gamma) + 
k^{2}h \rbrack /\alpha 
\end{equation}
Using Eq.(\ref{EQ.hy}) we find that to leading order at 
$\tau = \tau_{2-}$
\begin{equation} 
\Phi = \frac{p+1}{2p+1} \Sigma,\Gamma = \frac{p}{2p+1} \Sigma,
\end{equation}
\begin{equation} 
\Sigma \equiv M(p)y_{2}^{p}\sqrt{\frac{p}{p+1}}/a,
\label{eq.Sigma}
\end{equation} 
where the expansion parameter $y_{2}$ is given by 
\begin{equation}\label{eq.y2} 
y_{2} = k(\tau_{i} - \tau_{2}) = -pk/\alpha_{2} = -pk\tau_{2}
\end{equation}
noting that the continuity of $a$ and $\alpha \equiv a'/a$ is part of the 
matching conditions \cite{Der}.

Solving the simultaneous equations got by equating the different
expressions for the $\Phi,\Gamma$ pair at $\tau_{2-}$ and $\tau_{2+}$
gives
\begin{equation}\label{EQ.B+B-} 
B_{+} = -\epsilon_{2}B_{-} = 2a_{2}\Sigma,
\end{equation}
\begin{equation}\label{EQ.B-} 
B_{-} = 2p\sqrt{\frac{3p}{1+p}}M(p)y_{2}^{p-1},
\end{equation}
and the extreme smallness of $\epsilon_{2}$ means that we can put
$B_{+} = 0$ and thus neglect the cosine term in Eq.(\ref{EQ.nusoln})
for $\nu$. Thus Eqs.(\ref{EQ.nusoln},\ref{EQ.B-}) determine $\nu$ and
in consequence the gauge invariant amplitude $\Phi$,
Eq.(\ref{EQ.Phieps}), throughout the radiation era:
\begin{equation}\label{EQ.Phisoln} 
\Phi = B_{-}\lbrack\epsilon \cos (k(\tau - \tau_{2})/\sqrt{3}) 
-\sin (k(\tau - \tau_{2})/\sqrt{3})\rbrack/(a\epsilon^{2}).
\end{equation}

\subsection{Metric Tensor Perturbations}
\label{itre2}

For gravitons the matching is simpler \cite{Abb,Der}. All that is required 
is that $h_{ij}$ and its first time derivative be continuous. 

We have, in both the inflation and radiation eras,
Eq.(\ref{EQ.htilgrav}) for $\tilde{h}_{ij}$ with $\mu = \mu_{1}$ for
inflation, but in the radiation era given by $\mu'' + \mu(k^{2} -
a''/a) = \mu'' + k^{2}\mu = 0$ so that
\begin{equation}\label{EQ.mugrav2} 
\mu = G_{+}\cos(k(\tau - \tau_{2})) - G_{-}\sin(k(\tau - \tau_{2})).
\end{equation}   
Matching $(\mu/a)$ and $(\mu/a)'$ we find, to leading order in
$y_{2}$,
\begin{equation}\label{EQ.G+G-} 
G_{+} = 0,  G_{-} = pM(p)y_{2}^{p-1}.
\end{equation}
For power law inflation $p = -(1+\delta)$ where $0 < \delta \le
1/10$. We can make a direct comparison of Eq.(\ref{EQ.G+G-}) with
Eq.(\ref{EQ.B-}), which would suggest that the influence of density
perturbations on the CMB will be greater than that of gravity
waves. However there are more stages to go through.

\section{The Radiation to Matter Transition}
\label{RTM}

We shall assume that for $\tau > \tau_{1}$ the universe can be
characterized by matter with zero pressure for some value $\tau_{1}$ 
of the
conformal time. The transition time, $\tau_{tr}$, from the radiation
era, characterized by $p = \frac{1}{3}\rho$, can be as large as
$\sim(a_{1}H_{1})^{-1}$. Thus for waves influential in the CMBR,
section (\ref{infera1}), $k\tau_{tr}$ is not necessarily
small and then the instantaneous transition approximation is not
reliable. For the development of the matter components there have been
many detailed studies \footnote{See White et al.\cite{WSS}, Ma and
Bertschinger
\cite{MaB} and references therein.} using the Boltzmann equations for 
various components of the matter which might be present. These details
are important for the smaller scale structure of the CMBR, but not for
the larger scale structure. We wish to concentrate on this larger
scale in our comparison of the gravity wave and density perturbation
contributions, and thus to avoid questions of the components of
matter. Consequently we shall approximate the transition from
radiation, relativistic matter, to the non-relativistic matter
domination era by a smooth change in an overall equation of state.
The most significant parameter of such an approximation is the length
of time that the transition takes, and while we can indeed make a
reasonable estimate of this time, we shall also consider results as a
function of the transition time.

\subsection{Metric Scalar Perturbations}
\label{RTM1}

We postulate a smooth transition from the radiation era to the matter
era in which we first parametrize the density as a function of the
scale factor, $a$, by
\begin{equation}\label{EQ.roRTM} 
\rho = \varrho_{1}e^{-sr},
\end{equation}
where $s =s(r), r = \ln(a/a_{1}), a_{1} = a(\tau_{1}), \varrho_{1} =
constant$.  The energy conservation equation gives the pressure and
thus an implicit equation of state by
\begin{equation} 
\frac{d}{da}(\rho a^{3}) = -3pa^{2},
\end{equation}
\begin{equation}
p/\rho = \frac{1}{3}(r\frac{ds}{dr} + s)-1.\label{EQ.pRTM}
\end{equation} 
$s=4$ corresponds to the radiation era, $s=3$ to the matter dominated
era; we can postulate an explicit form for $s$ as a sufficiently
smooth and smoothly joining function of $r$ between those two
constants. The one we adopt is given in the appendix. Given such a
function we can find $\alpha \equiv r' = a'/a$ and the Hubble
parameter $H = r'/a$ as functions of $a$, and also, consequently,
$\tau = \tau(a)$ or equivalently $a = a(\tau)$, when given also the
values of $H = H_{1}, a = a_{1}$ at the beginning of the pure matter
era. Thus from the above equations:
\begin{equation} 
\varrho_{1} = 3 (H_{1}a_{1})^{2}/\kappa,
\end{equation}
\begin{equation}
\alpha = r' = H_{1}a_{1}\exp[(2-s)r/2].\label{eq.r's}
\end{equation}   
Adopting the notation that the transition begins at $a = a_{e}, r =
r_{e}, \tau = \tau_{e}$, we specify the length of the transition by  
\begin{equation}\label{EQ.rtrans} 
r_{trans} \equiv \frac{1}{2}(r_{1} - r_{e}) = 
\frac{1}{2}\ln(a_{1}/a_{e}).
\end{equation}
We now consider the density perturbations; these are given by the
development of $\nu$, defined in Eq.(\ref{EQ.nu}), or equivalently of
$u$\cite{Gri} where
\begin{equation}\label{EQ.unu} 
u = \frac{\alpha}{a}\nu = \alpha(\frac{dh}{dr} + \gamma h).
\end{equation}
In the radiation era $\nu$ took a simple sinusoidal form but now
evolves according to
\begin{eqnarray} 
&&\frac{d^{2}u}{dr^{2}} + 
\frac{du}{dr}\lbrack 3 + C + \frac{dB}{dr}\rbrack 
+u\lbrack (kc_{s}/\alpha)^{2}\nonumber\\ 
&&+ (1 + \frac{dB}{dr})(C + 1) +
2\frac{d^{2}B}{dr^{2}} + 2(\frac{dB}{dr})^{2}\rbrack = 0 .
\label{EQ.u}
\end{eqnarray}
where $c_{s}^{2} = p'_{0}/\rho'_{0}, B = (2-s)r/2, C = 3c_{s}^{2} -
 \frac{dc_{s}^{2}}{dr}/c_{s}^{2}$, with $p_{0}$ and $\rho_{0}$, the
 unperturbed pressure and density, given by
 Eqs.(\ref{EQ.roRTM},\ref{EQ.pRTM}). The initial conditions for
 solution are that $u,\frac{du}{dr}$ are continuous with the
 corresponding radiation era quantities; the equation has to be solved
 by numerical methods and the result is that the physical information
 on the development of the perturbations is passed continuously from
 the radiation era to the pure matter era. From the evolution of $u$
 by Eq.(\ref{EQ.u}), the evolution throughout the transition 
 of the gauge invariant metric scalar perturbation, $\Phi$, 
 can be computed using Eqs.(\ref{EQ.Phi},\ref{EQ.unu},\ref{EQ.hl}). 

In the matter era $a \propto \eta^{2}$ where
\begin{equation}\label{EQ.eta} 
\eta = \tau - \tau_{m} , \tau_{m} = constant,
\end{equation}
with the pressure being equal to zero. Consequently Eqs.(\ref{EQ.pe},
\ref{EQ.hl}) yield
\begin{equation} 
h  = C_{1} + C_{m}\eta^{-3},\label{EQ.hmat}\\
\end{equation}
\begin{equation}
h'_{l} = C_{1}k^{2}\eta/5 + C_{2}\eta_{1}^{3}/\eta^{4}
 + C_{m}k^{2}\eta^{-2}/2,\label{EQ.hlmat}
\end{equation}
where $C_{1}, C_{2}, C_{m}$ are constants; the synchronous gauge mode,
proportional to $C_{m}$, being non-physical it is just $C_{1},C_{2}$
that we require. The gauge invariant perturbation, $\Phi$, is given by
\begin{equation}\label{eq.Phimat} 
\Phi = h - \frac{\alpha}{k^{2}}h'_{l} = \frac{3}{5}C_{1} - 
\frac{2\eta_{1}^{3}}{k^{2}\eta^{5}} C_{2},
\end{equation}
and we note that $C_{m}$ does not appear. At the beginning of the pure 
matter era ($\tau = \tau_{1}, \eta = \eta_{1}, \alpha = \alpha_{1}$)
\begin{equation} 
\Phi(\tau_{1}) = \frac{3}{5}C_{1}-\frac{1}{2}(\alpha_{1}/k)^{2} C_{2},
\label{eq.Phim}
\end{equation}
\begin{equation}
\Phi'(\tau_{1}) = \frac{5}{4}\alpha_{1}(\alpha_{1}/k)^{2} C_{2}
\label{eq.Phi1}
\end{equation}
and the continuity of $\Phi,\Phi'$, calculated through the radiation
and transition eras, gives $C_{1},C_{2}$ in terms of (i) the
inflationary power $p$ and $\tau_{2}$ (or equivalently $a_{2}$) by
Eqs.(\ref{EQ.B-},\ref{EQ.Phisoln}) and of (ii) the parameters of the
radiation to matter transition.

We can also investigate what the sudden transition approximation 
gives. Analogously to section \ref{itre1} we then have to enforce  
the continuity of $\Phi,\Gamma$ at $\tau=\tau_{1}$; $\Gamma$ is given 
by Eq.(\ref{eq.Gamma}) where now $\epsilon \equiv k/\alpha\sqrt{3}$ 
is evaluated at $\tau=\tau_{1},\epsilon = \epsilon_{1}$. From this 
we find that if we denote $\Gamma_{1-},\Phi_{1-}$ to be the values 
at the radiation side of the interface then 
\begin{equation} 
C_{1} = \Phi_{1-}(1-\frac{2}{3} \epsilon_{1}^{2}) + \Gamma_{1-}
\label{eq.C1}
\end{equation}
\begin{equation} 
C_{2} =\frac{18}{5}\epsilon_{1}^{2} 
\{-\frac{2}{3}\Phi_{1-}(1+\epsilon_{1}^{2}) + \Gamma_{1-}\}
\label{eq.C2}
\end{equation}
Thus for the coefficient of the growing component we find
\begin{equation} 
C_{1} = \{B_{+}(2\cos y - \epsilon_{1}\sin y)-
B_{-}(2\sin y + \epsilon_{1}\cos y)\}/6a
\label{eq.C1B}
\end{equation}
where $y \equiv k(\tau_{1}-\tau_{2)})/\sqrt{3} \approx 
k\tau_{1}/\sqrt{3} =\epsilon_{1}$. Here $B_{+} = 0$ is, as we have 
found above in Eq.(\ref{EQ.B+B-}), a result good to many powers of  
10 relative to $B_{-}$. So we write 
\begin{equation} 
C_{1} = B_{-}(2\sin y + \epsilon_{1}\cos y)/6a
\label{eq.C1C}
\end{equation}
and in the discussion section we shall compare this with 
the corresponding result of ref.\cite{Gri}.

An essential difference between this radiation era to matter era 
transition and the inflation to radiation era transition is that now 
we cannot necessarily expect, for values of $k$ relevant to the CMBR 
fluctuations, that the sudden approximation is nearly exact. We  
investigate this quantitatively below, taking account of both the 
growing, $C_{1}$, and the decaying,$C_{2}$, mode.

\subsection{Metric Tensor Perturbations}
\label{RTM2}

Throughout all cosmic eras in the FRW universe, gravitons of
primordial origin are given by
Eqs.(\ref{EQ.mugrav},\ref{EQ.htilgrav}). Thus their development
depends only on the evolution of the scale factor of the universe
through the function $a''/a$. From the beginning of radiation to the
present time the gravitational mode function $\mu$, or equivalently
the Bogoliubov coefficients \cite{Gri2,MHM}, evolve smoothly by
Eq.(\ref{EQ.mugrav}) given any twice differentiable scale parameter
$a(\tau)$. In the radiation era we have Eq.(\ref{EQ.mugrav2}) for
$\mu$ while in the matter era where $a''/a = 2\eta^{-2}$ we express
the solution in terms of spherical Bessel functions:
\begin{equation}\label{EQ.mugravmat} 
\mu = \sqrt{\frac{\pi z}{2}}\lbrack G_{1}J_{\frac{3}{2}}(z) + 
G_{2}J_{-\frac{3}{2}}(z) \rbrack,
\end{equation}
where $z = k\eta$ and $G_{1},G_{2}$ are constants. In the radiation to
matter transition $\mu$ is found by computation using the postulated
smooth $a(\tau)$ (see Appendix) and thus $G_{1},G_{2}$ are
found in terms of the $G_{-}$ of Eq.(\ref{EQ.G+G-}) through the
continuity of $\mu,\mu'$.

\section{The Sachs-Wolfe Effect}
\label{SW}

Both the scalar and tensor perturbations give rise to perturbations in
the wavelength of the photons of the CMBR. The observation of apparent
temperature fluctuations, to which the perturbations would give rise,
is well established and continuing. Scalar perturbations are usually
considered to be dominant and we shall first deal with these.

To be appropriate for observations nearly in the rest frame of the
earth the calculation should be done in a comoving frame implying,
\cite{Gri}, $T_{0}^{i} = -a^{-2}h'Q_{,i} =0$, and from
Eq.(\ref{EQ.hmat}) this requires $C_{m} = 0$ in the matter era. A
synchronous coordinate system which is not comoving can be changed to
comoving while remaining synchronous. So we thus complete the
definition of our coordinate system, which is continuous from the
beginning of radiation to the present time. This makes no difference
to our results for $C_{1},C_{2}$ in terms of the $B_{+},B_{-}$
determined at the beginning of the radiation era, as can be seen from
the discussion at the end of \ref{RTM1}. In the matter era 
$h$ and $h_{l}$ are now given by Eqs.(\ref{EQ.hmat},\ref{EQ.hlmat}) 
with $C_{m} = 0$.

We consider reception of the CMBR at the present time,$\eta =
\eta_{0}$, and emission at $\eta = \eta_{E}$ where 
$\eta_{E} \ge \eta_{1}$ is a time within the matter era. (For the  
matter era we use the more convenient conformal time 
of Eq.(\ref{EQ.eta})). Then in the synchronous 
comoving coordinate system the CMBR fractional temperature
variation as a function of the direction of observation specified by
the unit vector $e^{i}$ is \cite{SW} in quantum mechanical form
\begin{eqnarray} 
&&\frac{\delta T}{T}({\bf e}) = \frac{1}{2}\int_{0}^{\omega_{E
}}\frac{d\tilde{h}_{ij}}{d\eta}e^{i}e^{j}d\omega 
\nonumber\\
&& = -\sqrt{\frac{G}{2\pi^{2}}}\int_{0}^{\omega_{E}}\lbrack\int
\frac{d^{3}k}{\sqrt{2k}}\lbrace c_{k}\frac{dh_{l}}{d\eta}
\frac{{\bf k.e}}{k^{2}}\exp(i{\bf k.x}) + c.c.\rbrace
\rbrack d\omega.
\label{EQ.deltaT}
\end{eqnarray}
In the above equation, $\omega = \eta_{0} - \eta$, the integration is
along the light path $x^{i} = e^{i}\omega$ and the quantum mechanical
 annihilation and creation operators, $c_{k}$ and $c_{k}^{\dag}$, are
the same as those in Eq.(\ref{EQ.mutil}). Only $h_{l}$ appears in the
mode function in Eq.(\ref{EQ.deltaT}) because $dh/d\eta = 0$.

The angular correlation function for different directions 
${\bf e_{1}}$ and ${\bf e_{2}}$ is 
 \begin{equation}\label{eq.K} 
K=\langle 0 \left\vert \frac{\delta T}{T}( {\bf e_{1}})
\frac{\delta T}{T}({\bf e_{2}}) \right\vert 0 \rangle ,
\end{equation} 
where $\left\vert 0 \right\rangle$ is the vacuum state, appropriate 
to the operators $c_{k}$, of the universe at the beginning 
of inflation.  We are working in the Heisenberg picture where the 
cosmic state vector is constant and the operators vary; in 
Eq.(\ref{EQ.deltaT}) the variation of the quantum operators is 
subsumed in the variation of the mode functions.

Grishchuk \cite{Gri} has shown that, with $\cos\delta = 
{\bf e_{1}}.{\bf e_{2}}$,
\begin{equation} 
K = K(\cos\delta) = \sum_{l=0}^{\infty}K_{l}P_{l}(\cos\delta),
\end{equation}
\begin{equation}
K_{l} = (2l+1)l_{Pl}^{2}\int_{0}^{\infty}\frac{dk}{2k}
\left\vert  
\int_{0}^{k\omega_{E}}dx\frac{dh_{l}}{d\eta}(\eta_{R}-\frac{x}{k})
L_{l}\right\vert^{2},
\end{equation}
\begin{equation}
L_{l} = \lbrace(\frac{l(l-1)}{x^{2}}-1)x^{-\frac{1}{2}}
J_{l+1/2}(x) + 2x^{-\frac{3}{2}}J_{l+3/2}(x)\rbrace.
\label{eq.Kl}
\end{equation}

We can express this with more information given as follows: We have
shown that the parameters $C_{1},C_{2}$ of $h_{l}$ are linear 
functions of $B_{+},B_{-}$, say
\begin{equation}\label{eq.CB} 
C_{1} = C_{1-}B_{-} + C_{1+}B_{+}; C_{2} = C_{2-}B_{-} + C_{2+}B_{+}.
\end{equation} 
That is, for example, $C_{1-}$ is the value of $C_{1}$ when calculated 
for $B_{-} = 1, B_{+} = 0$. We know from 
Eqs.(\ref{EQ.B+B-},\ref{EQ.B-}) the value of $B_{-}$, $B_{+}$ being 
nearly zero, so that
\begin{eqnarray} 
K_{l} & = & (2l+1)(\frac{l_{Pl}}{\tau_{2}})^{2}
(\frac{\tau_{2}}{\tau_{1}})^{2p}M^{2}(p)\frac{12p}{p+1}
\int_{0}^{\infty}\frac{dk}{2k}(-k\tau_{1}p)^{2p}\nonumber\\
 & \times & \left\vert \int_{0}^{x_{E}}dx
\lbrack\frac{1}{5}C_{1-}z^{2} + C_{2-}(\frac{z_{1}}{z})^{3}\rbrack
\frac{1}{z}L_{l}(x)\right\vert^{2},\label{EQ.Kl2}
\end{eqnarray}
where $z = k\eta, x = k\eta_{0} - k\eta = k\omega$.

A similar procedure can be carried through for the gravitons, having  
the mode function of Eq.(\ref{EQ.mugravmat}) in the matter era with
 $G_{1} = D_{1-}G_{-}+D_{1+}G_{+}; G_{2} = D_{2-}G_{-}+D_{2+}G_{+}$.
 Knowing the value of $G_{-}$, and that $G_{+}$ is approximately zero  
, from Eq.(\ref{EQ.G+G-}) we find \cite{Gri2}
\widetext
\begin{eqnarray} 
K_{l} & = & (2l+1)l(l+1)\lbrack l(l+1)-2\rbrack
(\frac{l_{Pl}}{\tau_{2}})^{2}(\frac{\tau_{2}}{\tau_{1}})^{2p}M^{2}(p)
\int_{0}^{\infty}\frac{dk}{2k}(-k\tau_{1}p)^{2p}\nonumber\\
& \times & \left\vert \int_{0}^{x_{E}}dx
\lbrack D_{1-}J_{5/2}(z) -D_{2-}J_{-5/2}(z) \rbrack
\sqrt{\frac{\pi z}{2}}\frac{1}{a(\eta)}
x^{-\frac{5}{2}}J_{l+1/2}(x)\right\vert^{2}.\label{EQ.Klg}
\end{eqnarray}
\narrowtext

\section{Results}
\label{results}

To produce numerical results we have to specify parameters giving (i) 
the length of the transition (ii) the end of the transition when the 
matter era begins (iii) the background radiation emission time for 
the Sachs-Wolfe calculation. We can conveniently specify using the 
scale factor, $a$, or the red shift, $z$, where $z+1 = a_{0}/a$; our 
convention is that $a_{0} \equiv a({\rm present}) = 1$ 

For (iii) it is rather natural to take the time of last scattering 
to be at the beginning of the matter era and this we shall do. We 
can make a remark from the literature supporting a not too great 
sensitivity to this assumption. If the emission time were taken 
to be within the radiation to matter transition then the correction 
from including the integrated Sachs-Wolfe effect is expected to be 
about 4\% in the quadrupole moment from scalar perturbations and 
less for other moments\cite{TW}. 

For the begining of the matter era (zero pressure) we consider  
two values, mainly $z_{1} = 10^{3}$ but also with some 
illustrations from $z_{1} = 10^{3.3}$ to assess the sensitivity 
of the results to the choice of $z_{1}$. As stated above $z_{1}$ 
is also taken to be the emmision red-shift in the Sachs-Wolfe 
calculation: 
\begin{equation}\label{eq.ze} 
z_{E} = z_{1}
\end{equation}
The parameter for the length of the transition, from $a = a_{e}$ 
to $a_{1}$, is given by $r_{trans}$ of Eq.(\ref{EQ.rtrans}) so 
that $r_{trans} \equiv \frac{1}{2}ln(a_{1}/a_{e}) = 
 \frac{1}{2}ln(z_{e}/z_{1})$. We shall present some results for 
a range of values of $r_{trans}$ including $r_{trans} = 0$, that 
being the sudden transition approximation which has often been 
used. We consider that a likely physical value is round about 
5 e-foldings:  
\begin{equation}\label{eq.rtrans2} 
r_{trans} = 2.5
\end{equation}
This can be very roughly estimated from a likely range for 
$z_{EQ}$, which can then be used to make an estimate from the 
known \cite{MFB} analytic two fluid model for the transition.

Our results of course also depend on $p$, the power of inflation ,
Eq.(\ref{EQ.ap}). We present results for a range of values of $p$, and
firstly for the ratio of the gravitational to the density perturbation
contributions to the multipoles. We emphasize that the ratio is
calculated with the same physical model for both contributions and
with no significant approximations.  For the ratio no other parameters
enter other than the ones just stated. But in the absolute magnitude
of each contribution, Eqs.(\ref{EQ.Kl2},\ref{EQ.Klg}),there is a
common factor
\begin{equation}\label{eq.F2} 
F_{2}=(\frac{l_{Pl}}{\tau_{2}})^{2}(\frac{\tau_{2}}{\eta_{1}})^{2p} 
\end{equation} 
containing the parameter $\tau_{2}$, the time when the radiation 
era begins. The choice of $\tau_{2}$ finally specifies the 
absolute magnitude of the multipoles. Within the model $p,\tau_{2}$
 are the important and significant unknown parameters; $r_{trans}$ 
and $z_{1}$ also have importance, as will be shown, but more is 
known about the possible values of these. 

As outlined in \ref{itre} our calculation of the radiation to 
matter transition is good for larger angular scale. There is an  
approximate relation between the correlation angle $\theta$ and 
the order of the multipole, $l$, which would give the main  
contribution to such a correlation \cite{LL}:
\begin{equation}\label{EQ.theta} 
\frac{\theta}{1^{o}} \approx \frac{60}{l}. 
\end{equation} 
We give results for multipoles with $l \leq 8$ corresponding by
Eq.(\ref{EQ.theta}) to angular scales $\theta \geq 8^{o}$. First 
we present the ratios, $(T/S)_{l}$, of the gravitational wave to 
the density wave Legendre series coefficients, $K_{l}$, of  
Eqs.(\ref{EQ.Kl2},\ref{EQ.Klg}); these ratios are independent of 
$\tau_{2}$. In Table 1 we show these for 4 different 
values of $p$ with $r_{trans} = 2.5$ and $z_{E} = z_{1} =  
10^{3}$.

From Table 2 we see,that for any particular multipole,  
the variation of $T/S$ with $p$ is approximately $(p+1)/p$. 
This is not quite unexpected as the ratio of gravitational to 
density wave amplitudes in the radiation era is given by 
Eqs.(\ref{EQ.B-},\ref{EQ.G+G-}) as $\sqrt{(p+1)/12p}$. However 
Table 2 shows that the ratio in observable multipoles, 
$(T/S)_{l}$, is bigger than $(p+1)/12p$ by a factor of order  
100.

Production of an adequate amount of inflation requires the cosmic  
time power $q \geq 10$. From Table 1 (and taking account of the   
proportionality to $(p+1)/p = 1/q$) we see that for $q < 100$ 
gravitational  waves make a non-negligable contribution and that for 
$q$ of order $10$ the contribution is similar to that of the density 
perturbations. Gravitational waves are proportionally most   
important in the quadrupole and octopole moments and Table 3  
 shows that these are the largest multipoles. 

Table 3 exhibits the absolute magnitude of the density  
 component of the multipoles, from which the graviton induced 
 component can be found using Table 1. Since $\tau_{1}$ 
 is approximately known the most important adjustable parameter 
 of the model here is $\tau_{2}$, the conformal time at the 
 beggining of the radiation era $\approx$ end of inflation. 
In the limit as $p \rightarrow -1$, $F_{2} \rightarrow 
(\frac{l_{Pl}\tau_{1}}{\tau_{2}^{2}})$.

We can find an order of magnitude of $\tau_{2}$, or equivalently
the more convention-free red-shift at the beginning of the 
radiation era, $z_{2}+1 = a_{0}/a_{2}$, by taking the estimate  
from observation of the quadrupole moment \cite{WSS,COBE}. If we insert  
$Q_{rms-PS} \approx 18 \mu K$ then with $p = -1.112,z_{1}=10^{3}$ 
we find,  where $h$ is the Hubble parameter 
\begin{equation}
z_{2} \approx  2.2\times10^{27}h^{-.47}, 
\end{equation}
and inserting  $6 \mu K$ gives 1.3 instead of 2.2 in the 
above \footnote{The amplitude $Q$ is related to the correlation 
multipole by $Q = T_{0}\sqrt{K_{2}}$ \cite{WSS}.}.

Table 4 shows the absolute values in units of $10^{4}F_{2}$ 
of the scalar and tensor quadrupoles for a range of values of 
the transition length $r_{trans}$ and $p = -1.112, q = 10$. 
The values for $r_{trans} = 0$ were obtained using the same  
sudden transition methods as those for inflation to 
radiation in \ref{itre}. For $r_{trans} > 0.4$ we 
use the smooth model transition of the Appendix. (For  
$0 < r_{trans} < 0.4$ the results are model dependent.) 
We note the approximate constancy of these results when the number 
of e-folds ($= 2r_{trans}$) is between 1 and 6 but that the sudden 
transition results are somewhat different, $T/S$ then being about 3/2  
larger. If we were to let the range of values of $k/\alpha$   
to be even smaller than those relevant for the CMBR observations 
we would expect these numbers to approach nearer to equality. 
Using the comoving gauge Lyth \cite{Lyt1} showed rather generally 
that, for small enough values of $k/\alpha$, the development of  
the density perturbation from one well-defined cosmic era to 
another is independent of the type of transition provided it be 
reasonably well behaved.

All the results discussed above were for values of the 
relevant red-shifts given by $z_{E} = z_{1} = 10^{3}$. 
If we increase this value to $10^{3.3}$ the scalar and 
tensor contributions to the multipole moments increase by  
a factor of around 4; an increase is expected as the Sachs-
Wolfe calculation is applied to a longer photon path. The 
ratio $(T/S)_{2}$, with $p=-1.122$, changes from 1.3 to  1.2 
with similar changes for the other multipoles.

\section{Discussion}
In this work we have used the same power law inflationary
model to evaluate and compare the density perturbation and
gravitational wave contributions to the CMBR fluctuations;
the transition from radiation domination to matter domination
has been appropriately treated as a gradual rather than a
sudden transition, but avoiding particle-spectrum dependent
details. This latter limits the precise validity of our
calculation to multipoles influencing correlation angles
greater than about 5 degrees. For power law inflation, 
$t^{q}$, the larger part of the quadrupole moment comes 
from gravitational waves if $q < 13.5$; the corresponding 
equality value for other multipoles is $\approx 10$. At 
$q=20$ gravitational waves contribute about $1/3$ to 
correlation Legendre coefficients of $l \leq 8$ and $10\%$ at 
$q \approx 100$.    

We followed the paper of Grishchuk \cite{Gri} by evolving using the  
metric perturbations, and also by making computations in the 
synchronous gauge, but our  
usage of this gauge, and our tracking of the perturbations from  
one cosmic era to the next, quite differ from his. In particular 
we made no significant use of non-physical synchronous 
gauge modes; we used gauge invariant variables in critical 
calculations of transitions and also, where a transition was
treated as sudden, we used the Lichnerowicz matching conditions 
\cite{Lic} as interpreted by Deruelle and Mukhanov \cite{Der}. 

The message delivered by our work on the relative importance of 
gravitational waves appears to differ from that of Grishchuk 
\cite{Gri}. Since the question can be raised as to where this 
difference comes from it may be of interest to identify the 
detailed reasons: (i) Firstly in ref.\cite{Gri} a special form for    
a continuous transition from inflation to radiation is used. In 
principle, with an appropriate choice of form, there should be 
nothing wrong with such a treatment; there is no interface and all 
quantities are continuous, this being indeed the principle we have 
used in the radiation era to matter era transition. Trouble seems 
to arise in ref.\cite{Gri} when the transition is taken to the zero 
transition time limit. This is particularly evident in the final 
treatment of $\gamma \equiv 1-\alpha'/\alpha^{2}$ which increases  
from the small value $(p+1)/p$ (in our notation) to 2 during the 
transition. The nature of the final expressions obtained in 
ref.\cite{Gri} for the equivalent of $B_{+}$ and $B_{-}$ and other 
quantities make it necessary, for a reasonable result, to rather  
arbitrarily put $\gamma = 2$ in these expressions \footnote{This 
non-rigorous treatment of $\gamma$ has also been criticised by Deruelle 
and Mukhanov \cite{Der}.}. The result we find is, on evaluation of 
Grishchuk's expression for the dominating constant which we call 
$B_{-}$, that it is smaller by a factor $\sqrt{(p+1)/p}$ than our 
expression given by Eqs.(\ref{eq.Sigma}),(\ref{EQ.B+B-}),
(\ref{EQ.B-}). This makes a factor of $(p+1)/p$ smaller for the 
density perturbation contribution to the final result. 
(ii) Secondly there is the question of the evaluation of the 
constants $C_{1}$ and $C_{2}$ of the matter era, which involves the 
transition from the radiation era. In expression eq.(82) of ref.
\cite{Gri} for $C_{1}$, this expression has been treated as 
sudden (analytic expressions such as those of ref.\cite{Gri} cannot 
otherwise be obtained); there are also other approximations 
equivalent to taking only the first term in eq.(\ref{eq.C1C}). 
Eq.(82)\cite{Gri} is {\it greater} by a factor $3/2$ than our first 
term in addition to having the {\it smaller} value of $B_{-}$ just 
related above. In our opinion this factor $3/2$ 
 arises from the non-trivial usage of 
the radiation gauge mode which is determined in ref.\cite{Gri} by 
an extra continuity condition not arising from Lichnerowicz matching. 
Thus we believe there are two parts of the density perturbation 
calculation of ref.\cite{Gri} which are incorrect.
                                  
A number of papers \cite{cncnss} by various authors found 
results on the relative importance of density and gravity wave 
perturbations from inflation in the CMBR fluctuations. These 
used particular properties of the various models in fairly 
complicated careful deductive processes. (For a recent account of a 
large range of density perturbation calculations, also referring to 
gravity wave contributions to the CMBR, and with  
complete references, see ref.\cite{Lyt2}.) By contrast we have used 
a direct calculation following in a unitary way 
both density and gravitational wave contributions from birth 
to observation, in a certain type of inflationary model.
And our conclusions are not in great disagreement with the general 
concensus, for power law inflation, of those previous papers.  

\section*{Acknowledgments}
We thank Andrew Liddle for a number of discussions and for valuable  
comments. L. E .M. acknowledges the support of a Praxis XXI grant, 
Lisbon.

\appendix
\section*{}
We adopt the following form for the function
$s(r), r = \ln(a/a_{1})$, of Section \ref{RTM1} where  
\begin{equation}\label{eq.a1} 
s(r_{e}) = 4 ,s(r_{1}) = 3 , 
\end{equation}
and $s(r)$ is specified between $r = r_{e}$ where the radiation 
era ends and  $r = r_{1}$ where the matter era begins. In the 
radiation and matter eras $\rho_{0} \propto \exp(-4r)$ and 
$\exp(-3r)$ respectively so the join with the transition era can 
be made arbitrarily smooth by specifying that  
\begin{equation}\label{eq.a2} 
 \frac{d^{m}s}{dr^{m}} = 0 ; r = r_{e}, r_{1} ,
\end{equation}
up to the necessary $m$. We take $m = 3$ and implement $s(r)$ by 
the polynomial form 
\begin{eqnarray}
s(r) & = & s_{0} + c(r-r_{0})\lbrack 1-x^{2}+\frac{3}{5}x^{4}-
\frac{1}{7}x^{6}\rbrack,\nonumber\\
r_{0} & = & (r_{e}+r_{1})/2 , s_{0} = (s_{e}+s_{1})/2 =7/2,\nonumber\\
c & = & 35(s_{1}-s_{e})/16(r_{1}-r_{e}),\nonumber\\
x & = & (r-r_{0})/(r_{1}-r_{0}).\label{a3} 
\end{eqnarray}
The degree of smoothness specified by \ref{eq.a2} with $m=3$, 
ensures the smooth joining at $r=r_{e}$ and $r_{1}$ of the 
coefficients.

\newpage
\begin{table}
\caption{The ratio $(T/S)_{l}$ of tensor to scalar multipoles of
 the correlation function, $K$,  for inflation $\propto t^{q}$, 
where $t$ is cosmic time; $q=p/(p+1)$.}
\label{T1} 
\begin{tabular}{|c|c|c|c|c|} 
$p+1$ & $-0.05$ & $-0.08$ & $-0.112$ & $-0.20$\\ \hline
$q$ & $21.0$ & $13.5$ & $9.9$ & $6.0$\\ \hline
$(T/S)_{2}$ & $0.64$ & $0.99$ & $1.33$ & $2.14$\\ 
$(T/S)_{3}$ & $0.52$ & $0.81$ & $1.09$ & $1.77$\\
$(T/S)_{4}$ & $0.41$ & $0.64$ & $0.88$ & $1.45$\\
$(T/S)_{5}$ & $0.46$ & $0.71$ & $0.97$ & $1.59$\\
$(T/S)_{6}$ & $0.44$ & $0.69$ & $0.95$ & $1.58$\\
$(T/S)_{7}$ & $0.39$ & $0.61$ & $0.84$ & $1.41$\\
$(T/S)_{8}$ & $0.43$ & $0.68$ & $0.93$ & $1.57$\\ 
\end{tabular}
\end{table}

\begin{table}
\caption{$\alpha_{l} \equiv (T/S)_{l}\frac{p}{p+1}$}
\label{T2} 
\begin{tabular}{|c|c|c|c|c|} 
$p+1$ & $-0.05$ & $-0.08$ & $-0.112$ & $-0.20$\\ \hline
$q$ & $21.0$ & $13.5$ & $9.9$ & $6.0$\\ \hline
$\alpha_{2}$ & $13.4$ & $13.4$ & $13.2$ & $12.8$\\ 
$\alpha_{3}$ & $10.9$ & $10.9$ & $10.8$ & $10.6$\\
$\alpha_{4}$ & $8.6$ & $8.6$ & $8.7$ & $8.7$\\
$\alpha_{5}$ & $9.7$ & $9.6$ & $9.6$ & $9.5$\\
$\alpha_{6}$ & $9.2$ & $9.3$ & $9.4$ & $9.5$\\
$\alpha_{7}$ & $8.2$ & $8.2$ & $8.3$ & $8.5$\\
$\alpha_{8}$ & $9.0$ & $9.2$ & $9.2$ & $9.4$\\ 
\end{tabular}
\end{table}

\begin{table}
\caption{Multipoles $S_{l}=K_{l}$ of density perturbations in units  
$10^{4}F_{2}$}   
\label{T3} 
\begin{tabular}{|c|c|c|c|c|} 
$p+1$ & $-0.05$ & $-0.08$ & $-0.112$ & $-0.20$\\ \hline
$q$ & $21.0$ & $13.5$ & $9.9$ & $6.0$\\ \hline
$S_{2}$ & $3.41$ & $2.48$ & $2.08$ & $1.82$\\ 
$S_{3}$ & $2.29$ & $1.62$ & $1.32$ & $1.07$\\
$S_{4}$ & $2.01$ & $1.40$ & $1.12$ & $0.86$\\
$S_{5}$ & $1.43$ & $0.98$ & $0.78$ & $0.58$\\
$S_{6}$ & $1.24$ & $0.84$ & $0.65$ & $0.47$\\
$S_{7}$ & $1.20$ & $0.80$ & $0.62$ & $0.43$\\
$S_{8}$ & $0.95$ & $0.63$ & $0.48$ & $0.33$\\ 
\end{tabular}
\end{table}

\begin{table}
\caption{Scalar and tensor quadrupoles (units $10^{4}F_{2}$), 
of the correlation function $K$,  
and their ratio as functions of the radiation to matter 
transition length}  
\label{T4} 
\begin{tabular}{|c|c|c|c|} 
$r_{trans}$ & $S_{2}$ & $T_{2}$ & $(T/S)_{2}$\\ \hline
$0$ & $1.72$ & $3.44$ & $2.00$\\ 
$0.5$ & $2.08$ & $2.78$ & $1.34$\\
$2.0$ & $2.12$ & $2.81$ & $1.33$\\
$2.5$ & $2.08$ & $2.77$ & $1.33$\\
$3.0$ & $2.09$ & $2.73$ & $1.31$\\ 
\end{tabular}
\end{table}
\end{document}